\documentstyle[preprint,aps,psfig]{revtex}
\begin{document}
\title{ A Note on the Positive Constant Curvature Space}

\author{Rong-Gen Cai\footnote{Email address: cairg@itp.ac.cn}}

\address{Institute of Theoretical Physics, Chinese Academy of Sciences,
   \\
 P.O. Box 2735, Beijing 100080, China}

\maketitle
\begin{abstract}
We construct  a positive constant curvature space by identifying
some points along a Killing vector in a de Sitter Space. This
space is the counterpart of the three-dimensional Schwarzschild-de
Sitter solution in higher dimensions. This space has a
cosmological event horizon,  and is of the topology ${\cal
M}_{D-1}\times S^1$, where ${\cal M}_{D-1}$ denotes a
$(D-1)$-dimensional conformal Minkowski spacetime.

\end{abstract}

\newpage

 In recent years people have been paying particular
interest on the constant curvature spacetimes. For the case of
negative constant curvature space, namely the anti-de Sittter
(AdS) space, it should be  mainly attributed to the celebrated
AdS/CFT (conformal field theory) correspondence~\cite{AdS}, which
states that the string or M theory on an AdS space times a compact
space is dual to a strong coupling conformal field theory residing
on the boundary of the AdS. For the case of positive constant
curvature space, namely the de Sitter (dS) space, one of the
reasons responsible for the particular interest is that defined in
a manner analogous to the AdS/CFT correspondence, an interesting
proposal, which says that there is a dual between quantum gravity
on a dS space and a Euclidean CFT on a boundary of the dS space,
the so-called dS/CFT correspondence, has been suggested
recently~\cite{Strom}. Another reason is that our universe may be
an asympotically dS spacetime~\cite{Cos}. However, different from
the cases for the asymptotically flat or AdS spacetimes, we have
not yet a fundamental description of asymptotically dS spacetimes
in the sense of quantum gravity: there is presently no fully
satisfactory embedding of dS space into string theory~\cite{BDM}.

In $D(\ge 3)$ dimensions, the AdS space has the topology $S_1
\times R^{D-1}$. By identifying  points in the universal covering
space of the AdS space in a some manner, one can construct various
spacetimes which are locally equivalent to the AdS space. It is
well-known that the BTZ black hole~\cite{BTZ} just belongs to this
kind of spacetime, which is  constructed by identifying points
along a boost Killing vector in a three dimensional AdS space. Its
counterparts are discussed in four dimensions~\cite{Amin} and in
higher dimensions~\cite{Bana}, respectively. The topology
structure of the BTZ black hole is ${\cal M}_2 \times S^1$, where
${\cal M}_n$ denotes a conformal Minkowski spacetime in $n$
dimensions. So the topology of the BTZ black hole is not peculiar,
but its higher dimensional counterparts have the topology ${\cal
M}_{D-1} \times S^1$ in the case of $D$ dimensions, which is quite
different from the usual one ${\cal M}_2 \times S^{D-2}$. These
black holes look strange in the sense that the exterior of black
holes is time-dependent and there is no globally defined timelike
Killing vector in the geometry of these black holes~\cite{Holst}.
So it is quite difficult to discuss the thermodynamics associated
with the black hole. In spite of the peculiar feature of the black
hole geometry, in \cite{Cai} we applied the surface counterterm
approach to a five dimensional constant curvature black hole, and
obtained the stress-energy tensor of dual conformal field theory.
 For other methods to study the constant curvature black holes,
see~\cite{Bana,Mann}.

On the other hand,  a $D(\ge 3)$-dimensional dS space has the
topology $R_1\times S^{D-1}$. Different from the AdS case, the dS
space has a cosmological horizon and associated Hawking
temperature and thermodynamic entropy $S$~\cite{Gibbons}:
\begin{equation}
\label{1eq1}
 S=\frac{\Omega_{D-2}}{4G}l^{D-2},
 \end{equation}
 where $G$ is the gravitational constant in $D$ dimensions, $l$ is
 the radius of dS space and $\Omega_{D-2}$ denotes the volume of a
 $(D-2)$-dimensional unit sphere. Due to this,  it is argued that
 the dS space should be described by a theory with a finite number
 $e^S$ of independent quantum states; the cosmological constant
 $\Lambda$
 should be understood as a direct consequence of the finite
 number of states $e^N$ in the Hilbert space describing the
 world~\cite{Banks,Bousso,Fischler}. Thus a natural consequence is that
 a necessary condition for a (coarse-gained) spacetime to be
 described by a theory with $e^N$ states is that the entropy
 accessible to any observer in that spacetime must
 obey~\cite{BDM}
 \begin{equation}
 S\le N.
 \end{equation}
 According to the ``$\Lambda \sim N $ correspondence"~\cite{Bousso},
 the theory with $e^N$ states may describe the set
 {\bf  all}$(\Lambda(N))$ of spacetimes: the class of spacetimes with
 positive cosmological constant $\Lambda(N)$, irrespective of
 asymptotic conditions and types of the matter
 present~\cite{BDM}. Applying the condition (\ref{1eq1}) to the
 candidate set {\bf all}$(\Lambda(N))$, a suggestion named ``N bound"
 has been proposed in~\cite{Bousso}, which in four dimensions states:
 {\it In any universe with a positive cosmological constant
 $\Lambda$ (as well as arbitrary additional matter that may well
 dominate at all times) the observable entropy $S$ is bounded by
 $N=3\pi/G\Lambda$}. Roughly specking, the ``N bound" means that
 the observable entropy in the
 spacetimes with a positive cosmological constant must be less than
 or equal to the entropy (\ref{1eq1}).  This ``N bound" leads naturally to the
 Bekenstein entropy bound of matter in dS spaces~\cite{Bousso2}.
 The geometry of the dS space looks simple, but this space has
 much non-trivial physics.

Similar to the negative constant curvature case, in this note we
construct a positive constant curvature spacetime by identifying
points along a rotation Killing vector in a dS space. Our
solutions turns out to be counterparts of the three-dimensional
Schwarzschild-dS solution in higher dimensions, and have an
associated cosmological horizon. There is a parameter in the
solution, which can be explained as the size of cosmological
horizon.

Let us begin with the five dimensional case.  A five dimensional
dS space can be understood as a hypersurface embedded into a six
dimensional flat space,  satisfying
\begin{equation}
\label{2eq1}
 -x_0^2+x_1^2 +x_2^2 +x_3^2 +x_4^2 +x_5^2 =l^2,
 \end{equation}
 where $l$ is the radius of the dS space. This dS space has
 fifteen Killing vectors, five boosts and ten rotations.
 Consider a rotation Killing vector $\xi =(r_+/l)(x_4\partial _5
 -x_5\partial_4)$ with norm $\xi^2 = r_+^2/l^2 (x_4^2 +x_5^2)$,
 where $r_+$ is an arbitrary real constant. The norm is always
 positive or zero. In terms of the norm,  the hypersurface
 (\ref{2eq1}) can be expressed as
 \begin{equation}
 \label{2eq2}
 x_0^2 =x_1^2 +x_2^2 +x_3^2 +l^2 (\xi^2/r_+^2-1).
 \end{equation}
 From (\ref{2eq2}) we see that when $\xi^2=r_+^2$, the hypersurface
 reduces to a null one
 \begin{equation}
 \label{2eq3}
 x_0^2=x_1^2 +x_2^2 +x_3^2,
 \end{equation}
 while $\xi^2 =0$, it goes to a hyperboloid
 \begin{equation}
 \label{2eq4}
 x_0^2 =x_1^2 +x_2^2 +x_3^2 -l^2.
 \end{equation}
 The null surface (\ref{2eq3}) has two pointwise connected
 brances, called $H_f$ and $H_p$, described by
 \begin{eqnarray}
 H_f:&& x_0=+\sqrt{x_1^2 +x_2^2 +x_3^2}, \nonumber \\
 H_p:&& x_0=-\sqrt{x_1^2 +x_2^2 +x_3^2}.
 \end{eqnarray}
 The hyperboloid (\ref{2eq4}) has two connected surfaces,
 named $S_f$ and $S_p$, defined as
 \begin{eqnarray}
 S_f:&& x_0 =+\sqrt{x_1^2 +x_2^2 +x_3^2-l^2}, \nonumber \\
 S_p:&& x_0 =-\sqrt{x_1^2 +x_2^2 +x_3^2-l^2}.
 \end{eqnarray}
 We plot the dS space in Fig.~1. Each point in the figure
 represent a pair $x_4$ and $x_5$ with $x_4^2 +x_5^2$ fixed.
 On the surfaces $S_f$ and $S_p$, we have $\xi ^2=0$, while
 $\xi^2 =r_+^2$ on the surfaces $H_f$ and $H_p$ of the two
 cones. In the region between $S$ and $H$, one has $0<\xi^2<r_+^2$,
 and  $\xi^2 >r_+^2 $ inside the two cones. Thus an observer located
 at the surface $S_f$ or $S_p$ cannot see events which happen
 inside the cone with null surface $H_f$ or $H_p$. That is, there
 does not  exist any communication between the inside the cones
 and outside the cones. Therefore, the surface $H_f$ can be regarded as
  the future cosmological event horizon and $H_p$ as the past one.

Identifying points along the Killing vector $\xi$, another
one-dimensional manifold becomes compact and isomorphic to $S^1$.
Thus we finally obtain a spacetime with cosmological horizon and
with the topology ${\cal M}_4\times S^1$. The Penrose diagram of
the positive constant curvature spacetime is plotted in Fig.~2.
Similiar to the case of negative constant curvature~\cite{Bana},
we may describe the spacetime in the region with $0\le \xi^2 \le
r_+^2$ by introducing six dimensionless local coordinates $(y_i,
\phi)$,
\begin{eqnarray}
\label{2eq7}
 && x_i=\frac{2ly_i}{1+y^2}, \ \ \ i=0,1,2,3 \nonumber \\
 && x_4=\frac{lr}{r_+}\sin\left(\frac{r_+\phi}{l}\right),
 \nonumber \\
 && x_5=\frac{lr}{r_+}\cos\left(\frac{r_+\phi}{l}\right),
 \end{eqnarray}
 where
 \begin{equation}
 r=r_+\frac{1-y^2}{1+y^2}, \ \ \ y^2=-y_0^2+y_1^2 +y_2^2+y_3^2.
 \end{equation}
Here the coordinate range is $-\infty <y_i <+\infty$, and $-\infty
<\phi <+\infty$ with the restriction $-1<y^2<1$ in order to keep
$r$ positive. In coordinates (\ref{2eq7}), the induced metric is
\begin{equation}
\label{2eq9}
 ds^2=\frac{l^2(r+r_+)^2}{r_+^2}(-dy_0^2+dy_1^2+dy_2^2+dy_3^2)
 +r^2 d\phi^2,
 \end{equation}
 which has the same form as the case of negative constant
 curvature~\cite{Bana}. However, it should be pointed out here
 that the coordinates (\ref{2eq7}) and the definition of $r$ are
 different from the corresponding ones in the constant curvature
 black holes.
In these coordinates it is evident that the Killing vector is
  $\xi =\partial_{\phi}$ with norm $\xi^2=r^2$. Thus in these
  coordinates one has $r=0$ on the surfaces $S_f$ and $S_p$
  and $r=r_+$ on the horizons $H_f$ and $H_p$.  With the
  identification $\phi \sim \phi +2\pi n$, the solution has
  obviously the topology ${\cal M}_4 \times S^1$.

 It is trivial to generalize the five dimensional case to other dimensions.
  For the case of an arbitrary dimensions ($D \ge 3$), the dS space
is a hypersurface satisfying
\begin{equation}
-x_0^2 +x_1^2 +\cdots +x_{D-1}^2 +x_D^2=l^2,
\end{equation}
in a $(D+1)$-dimensional flat spacetime. Consider a rotation
Killing vector $\xi=r_+/l(x_D\partial_{D-1} -x_{D-1}\partial_{D})$
with norm $\xi^2=r_+^2/l^2(x^2_{D-1}+x^2_D)$ and identify points
along the orbit of this Killing vector, we can obtain a positive
constant curvature space with the topology ${\cal M}_{D-1}\times
S^1$. Introducing $(D+1)$ dimensionless coordinates like
(\ref{2eq7}), one has the induced metric
\begin{equation}
ds^2 =\frac{l^2(r+r_+)^2}{r_+^2}\eta_{ij}dy^idy^j +r^2d\phi^2,
\end{equation}
where $\eta_{ij}={\rm diag}(-1,1,\cdots,1)$ and $i,j=0, 1, \cdots,
D-2$.

Like the dS space, we can also introduce Schwarzschild coordinates
to describe the solution. Using local ``spherical" coordinates
$(t,r,\theta,\chi)$ defined as
\begin{eqnarray}
\label{2eq12}
 && y_0=f\cos\theta\sinh(r_+t/l), \ ~~~~~ y_1=f\cos\theta
 \cosh(r_+t/l), \nonumber \\
 && y_2=f\sin\theta \sin\chi, \ ~~~~~ y_3=f\sin\theta \cos\chi,
 \end{eqnarray}
 where $f=[(r_+-r)/(r+r_+)]^{1/2}$, and the coordinate range is
 $ 0< \theta <\pi/2$, $0<\chi <2\pi$ and $0<r<r_+$, we find that
 the solution can be expressed as
 \begin{equation}
 \label{2eq13}
 ds^2 =l^2 N^2d\Omega_3 +N^{-2}dr^2 +r^2d\phi^2,
 \end{equation}
 where $N^2=(r_+^2-r^2)/l^2$ and
 \begin{equation}
 \label{2eq14}
 d\Omega_3= -\sin^2\theta dt^2 +\frac{l^2}{r_+^2}(d\theta^2
 +\cos^2\theta d\chi^2).
 \end{equation}
 In these coordinates $r=r_+$ is the cosmological horizon.
 This solution is just the counterpart of a five dimensional constant
 curvature black hole in the Schwarzschild coordinates~\cite{Bana}.
 The only difference is that $N^2=(r^2-r_+^2)/l^2$ there is
 replaced by $N^2=(r_+^2-r^2)/l^2$ here.  In three dimensions, the
 corresponding induced metric is
 \begin{equation}
 \label{2eq15}
 ds^2=-(r_+^2-r^2)dt^2 +\frac{l^2}{r_+^2-r^2}dr^2 +r^2d\phi^2,
 \end{equation}
 After a suitable rescaling of coordinates it can be transformed
 to usually three-dimensional Schwarzschild-de Sitter
 solution~\cite{Park}.
 In (\ref{2eq14}), if $\theta=\pi/2$, it also reduces to the
 three-dimensional Schwarzschild-dS solution.

 Within the cosmological horizon  dS space is time-independent in the static
 coordinates.  The solution (\ref{2eq13}) looks also static, but it does not
 cover the whole region within the cosmological horizon, which can be seen from
 the definition of coordinates (\ref{2eq12}) because they must obey the
 constraint: $y_1^2-y_0^2=f^2\cos^2\theta \ge 0$,

As the black hole case, there is a set of coordinates which cover
the whole region within the cosmological horizon. They
are~\cite{Cai}
\begin{eqnarray}
\label{2eq16}
 && y_0=f\sinh(r_+t/l), \ ~~~~~~ y_1=f\cos\theta
\cosh(r_+t/l),
\nonumber \\
&& y_2=f\sin\theta \cos\chi \cosh(r_+t/l), \ ~~~~~~
y_3=f\sin\theta \sin\chi \cosh(r_+t/l).
\end{eqnarray}
In terms of these coordinates, the solution is described by
\begin{equation}
\label{2eq17}
 ds^2=l^2N^2\tilde {d\Omega_3} +N^{-2}dr^2 +r^2d\phi^2,
 \end{equation}
 where $N^2=(r_+^2-r^2)/l^2$ and
 \begin{equation}
 \label{2eq18}
 \tilde {d\Omega_3}=-dt^2
 +\frac{l^2}{r_+^2}\cosh^2(r_+t/l)(d\theta^2+\sin^2\theta
 d\chi^2).
\end{equation}
It can be seen from (\ref{2eq18})  that when $\theta =\chi =0$,
the solution will reduce to the three-dimensional Schwarzschild-dS
solution (\ref{2eq15}). Therefore the solution we constructed here
is the counterpart of the three-dimensional Schwarzschild-dS
spacetime in five dimensions.  Further it should be emphasized
here that both sets of coordinates (\ref{2eq12}) and (\ref{2eq16})
can be used within the cosmological horizon only. This can be seen
from the metrics (\ref{2eq13}) and (\ref{2eq17}): beyond the
horizon the signature of the solution changes. It is certainly of
interest to find a set of coordinates describing the outer region
of cosmological horizon.

In $D$ dimensions, the solution is
\begin{equation}
ds^2=l^2N^2d\tilde \Omega_{D-2} +N^{-2}dr^2 +r^2 d\phi^2,
\end{equation}
where $N^2$ is still the one given before, and
\begin{equation}
d\tilde\Omega_{D-2}=-dt^2
+\frac{l^2}{r_+^2}\cosh^2(r_+t/l)d\Omega_{D-3}.
\end{equation}
Here $d\Omega_{D-3}$ denotes the line element of a
$(D-3)$-dimensional unit sphere.

The Euclidean sector of the solution can be obtained via the
transformation, $t\to -i (\tau +\pi l/2r_+)$, in the solution
(\ref{2eq17}). In that case,  the line element (\ref{2eq18}) is
changed to
\begin{equation}
\label{2eq19}
  d\tilde \Omega_3 =d\tau^2 +\frac{l^2}{r_+^2}
      \sin^2(r_+\tau/l)(d\theta^2 +\sin^2\theta d\chi^2).
\end{equation}
In order  the $d\tilde \Omega_3$ to describe a regular
three-sphere, the $\tau$ must have a  range $0 \le \tau \le \beta$
with
\begin{equation}
\label{2eq20}
 \beta =\pi l/r_+,
\end{equation}

In the coordinates (\ref{2eq12}), the Euclidean sector of the
 solution is (\ref{2eq13}) with
 \begin{equation}
 d\Omega_3= \sin^2\theta d\tau^2 +\frac{l^2}{r_+^2}(d\theta^2
 +\cos^2\theta d\chi^2),
 \end{equation}
here $\tau$ has the range $0 \le \tau \le \tilde \beta $ with
\begin{equation}
\tilde \beta =2 \pi l/r_+,
\end{equation}
which differs from the value of $\beta$ (\ref{2eq20}). Although
the solution expressed in terms of the coordinates (\ref{2eq12})
does not cover the whole interior of cosmological horizon, the
Euclidean sector of solution is complete in both sets of
coordinates (\ref{2eq12}) and (\ref{2eq16}). Further, regarding
$\beta$ or $\tilde \beta$ as the inverse temperature of the
cosmological horizon seems problematic since calculating the
surface gravity $\kappa$ of the cosmological horizon shows it does
not satisfy the usual thermodynamic relation $\beta =2\pi
/\kappa$. Obviously, it is of great interest to discuss the
thermodynamics associated with the cosmological horizon in the
constructed spacetime.

Next we further consider the Euclidean solution in the coordinates
(\ref{2eq19}). Making a coordinate transformation
$r^2=r_+^2(1-R^2/l^2)$ and $\tau =l\varphi /r_+$, one can find
that the Euclidean solution (\ref{2eq17}) with (\ref{2eq19}) can
be expressed as
\begin{equation}
\label{e27}
 ds^2=\left(1-\frac{R^2}{l^2}\right) d(r_+\phi)^2
+\left(1-\frac{R^2}{l^2}\right)^{-1}dR^2 +R^2(d\varphi^2
  +\sin^2\varphi (d\theta^2 +\sin^2\theta d\chi^2)).
  \end{equation}
This is evidently the Euclidean solution of de Sitter space in the
static coordinates with the Euclidean time $r_+\phi$. Since the
$\phi$ has the period $2\pi$, so the solution (\ref{e27}) does not
describe a regular instanton. The regular instanton requires the
Euclidean time $r_+\phi$ has a period $2\pi l$. This implies that
the solution (\ref{2eq17}) has a conical singularity along the
circle $\phi$ if $r_+\ne l$ with a deficit angle $2\pi (1-r_+/l)$.
When $r_+ <l$, the deficit angle is less than $2\pi$, otherwise it
becomes negative. When $r_+=l$, the solution is regular without
any singularity.  This situation is the same as the case of three
dimensional Schwarzschild-dS solution~\cite{Park}.

In summary we have constructed a positive constant curvature space
by identifying points along a rotation Killing vector in a dS
space. This space is the counterpart of the three-dimensional
Schwarzschild-dS solution in higher dimensions. Also this space
can be viewed as corresponding counterpart of the negative
constant curvature black hole constructed in~\cite{Bana}. Unlike
the dS space, the positive constant curvature space constructed in
this note has the topology ${\cal M}_{D-1}\times S^1$. As the dS
space, however, it still has a cosmological horizon $r_+$. The
solution has a conical singularity along the circle $\phi$ if
$r_+\ne l$. It should be interesting to discuss the thermodynamics
of cosmological horizon and dual conformal field theory in the
spirit of dS/CFT correspondence.

\section*{Acknowledgments}
The author thanks H.Y. Guo and J.X. Lu for useful discussions, and
C.T. Shi for help in drawing the figures. This work was supported
in part by a grant from Chinese Academy of Sciences.

\begin{figure}
\psfig{file=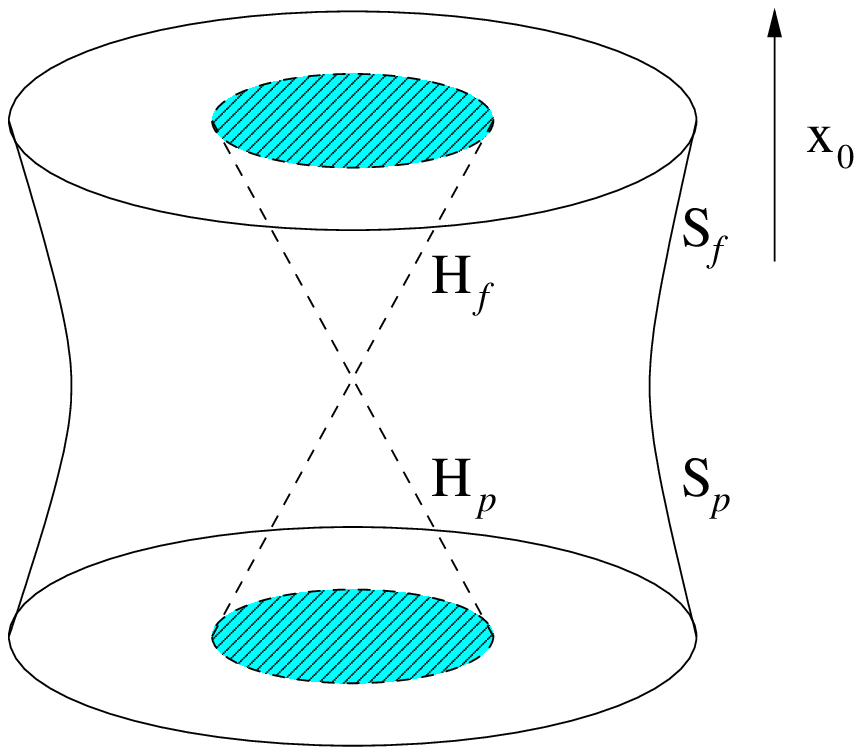,height=85mm,width=120mm,angle=0}
 \caption{de Sitter Space. Each point in the figure represent a pair
 of $x_4$ and $x_5$ with $x_4^2 + x_5^2$ fixed.}
\end{figure}
\vspace*{1.cm}
\begin{figure}
\psfig{file=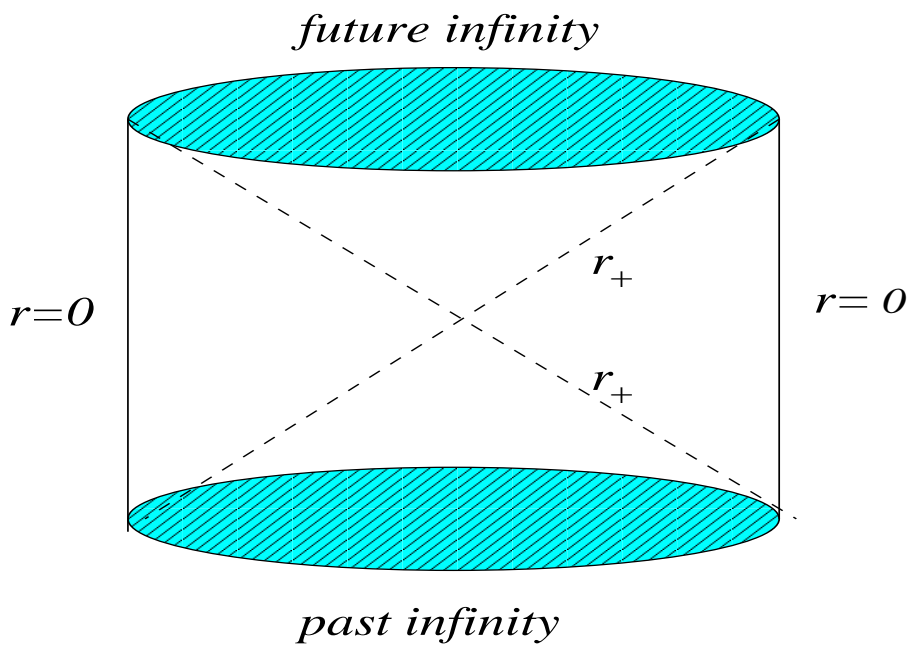,height=85mm,width=120mm,angle=0}
 \caption{The Penrose diagram for the positive constant curvature space.}
\end{figure}

\end{document}